\documentclass[aps,prl,twocolumn,showpacs,showkeys]{revtex4-1}

\begin{document}


\title{Preferential attachment alone is not sufficient to generate scale free random networks}


\author{Vijay K. Samalam}
\affiliation{Janelia Farm Research Campus, Howard Hughes Medical Institute, 19700 Helix Drive, Ashburn, VA 20147}


\date{\today}

\begin{abstract}
Many networks exhibit scale free behaviour where their degree distribution obeys a power law for large vertex degrees.  Models constructed to explain this phenomena have relied on preferential attachment where the networks grow by the addition of both vertices and edges, and the edges attach themselves to a vertex with a probability proportional to its degree.  Simulations hint, though not conclusively, that both growth and preferential attachment are necessary for scale free behaviour. We derive analytic expressions for degree distributions  for networks that grow by the addition of edges to a fixed number of vertices, based on both linear and non-linear preferential attachment, and show that they fall off exponentially as would be expected for purely random networks. From this we conclude that preferential attachment alone might be necessary but is certainly not a sufficient condition for generating scale free networks.  
\end{abstract}


\maketitle

Many natural and man made networks possess a scale free degree distribution where the tail of the distribution follows a power law \cite{SIAMReview:2009}. These networks are particularly interesting because they hint at complicated processes underlying their generation - processes sufficiently complex that the resulting networks cannot be described by a few simple parameters like the mean and standard deviation. Generative models that have been constructed to explain this behaviour assume networks grow by adding vertices one by one, with each vertex having a fixed number of either directed or undirected edges that connect to other existing vertices with a probability proportional to their degree \cite{Price:1976,Barabasi:1999}.  For most models, in the limit of large network size, the distribution for large degrees falls off with an exponent that is between 2 and 3.  The basic linear preferential model has been extended to include the addition and removal of edges after the vertices have been added and again the networks have been shown to be scale free \cite{Albert:2000,Dorogovtsev:2000,Krapivsky:2001,Moore:2006}.  Finally it has been shown that models, where the probability of attachment to a vertex goes as a power of the degree of the vertex, generate stretched exponentials when that power is less than one \cite{Krapivsky:2000}. For reviews of these and other results on random graphs in general see \cite{Albert:2002}.  For a particularly readable and pedagogical treatment of these results see \cite{newmanbook}.
\par
All these models are variants of the basic citation network, where the network grows both by the addition of vertices and of edges.   One of the characteristics of these models is that older vertices in the network - those added earlier in the growth process - have more time to acquire edges than more recent vertices and, in fact, it has been shown that these vertices on the average have a higher degree \cite{DMS:2000}. Simulations seem to indicate that both growth (of vertices) and preferential attachment are necessary for scale free behaviour although the result is not conclusive.  Detecting power law behaviours in simulation data or real world data is difficult because of the large fluctuations that occur in the tail of the distribution \cite{SIAMReview:2009}.  While simulation of the growth of a citation network without preferential attachment clearly shows exponential behaviour, simulation results for edges attached to a fixed number of vertices through preferential attachment never seem to have attained a stationary state - initially the degree distribution seems to show scale free behaviour but then the network reaches saturation, where all the vertices are fully interconnected \cite{Barabasi:1999}. In this paper we derive analytic expressions for the degree distribution of networks where edges attach themselves to a fixed number of vertices through preferential attachment (both linear and non-linear) and show that the degree distributions fall off exponentially just as you would expect for purely random networks.  Thus preferential attachment might be a necessary but is certainly not a sufficient condition for generating networks with scale free degree distributions.
\par
We construct our generative network model by considering $ n $ vertices and then grow the network by attaching edges to vertices, or more accurately ends of edges (stubs) to vertices.  At each time step we add the end of an undirected edge to one of the n vertices based on some probability $ p_{k} $.  The other end of the edge is then attached to any of the n vertices with a probability drawn from the same distribution $ p_{k} $.  At time t, (t even), the stubs attached to the vertices can be connected randomly to produce one, among many equally likely, network configurations all with the same degree  sequence and distribution.  In the process of growing, at any given time $ t $, the network will have $ t/2$ edges. At any time in the growth of the network, $ t = \sum\limits_{i=1}^n k_{i} $ where $ k_{i} $ is the degree of the $ i_{th} $ vertex. In this model the other end of the edge can be attached to the same vertex producing self-edges or ``tadpoles''.  We also do not rule out multi-edges, where two vertices can have more than one edge (''melons'').  It has been shown that the number of either self-edges or multi-edges remains constant as the network grows if the second moment of the degree distribution, $ \bar{k^2} = \sum\limits_{i=1}^{n} k_{i}^2/n $, is finite and independent of n. We will show that in our model the distribution falls off exponentially for large degrees and therefore in the large $ n $ limit the presence of self-edges or multi-edges should not have any bearing on our fundamental result.
\par
We derive a master equation for the evolution of the degree distribution by considering an ensemble of networks growing in the manner described above. At $t=0$, the network starts off with $ n $ vertices, and no edges. At every time step, we add a stub, or the end of an edge to a vertex with probability given by
\begin{equation}
p_k\left( \gamma\right)  \propto\left( a+k^\gamma\right) .
\end{equation} 
where a is a positive constant (typically set to around 1) which ensures that all vertices, including those with degree 0 get a base chance at acquiring edges, and $ \gamma $ is a positive constant between 0 and 1. (It has been shown that for $ \gamma > 1 $, a single vertex gains a non-zero fraction of all edges and we will not be considering it here \cite{Krapivsky:2000}.)
The properly normalized probability for the attachment of a stub to a vertex is given by
\begin{equation}\label{probk>1}
p_k\left( \gamma\right)   = \frac{N\left( k,t\right)  \left( a + k^\gamma\right)  }{n\left[ a + <k^\gamma>\left( t\right)  \right]},\quad k \ge 1 \qquad\mbox{and}
\end{equation}

\begin{equation}\label{probk0}
p_k\left(\gamma\right)   = \frac{N\left( k,t\right)  \left( a + \delta_{\gamma,0}\right)  }{n\left[ a + <k^{\gamma}>\left( t\right) \right]},\quad k=0,
\end{equation}
where $ N\left( k,t\right)  $ is the number of vertices of degree $ k $ at time $ t $.
In the above equations $ <k^\gamma>\left( t\right) =  \sum\limits_{i=1}^n k_{i}^\gamma/n$. In the linear attachment model, $ \gamma = 1 $, and 
\begin{equation}
<k^\gamma>\left( t\right)  = \frac{t}{n} = \bar{k},
\end{equation}
where $ \bar{k} $ is the mean degree of a vertex at time $ t $ in the evolution of the network.
Notice that when $ \gamma = 0 $ or when $ a \longrightarrow \infty $, $ p_{k} = N\left( k,t\right) /n$, where the stubs are attached to vertices perfectly randomly and the network generated will be a completely random graph. (The slightly different form for the probability for $ k = 0 $ ensures that it has the right limit when $ \gamma = 0 $ 
\par
With the degree distribution given by $ P\left( k,t\right) = N\left( k,t\right) /n$, the master equation for the time evolution of $ P\left( k,t\right) $ can be shown to be 
\begin{widetext}
\begin{equation}
P\left( k,t+1,\gamma\right) = P\left( k,t,\gamma\right) \left\lbrace 1 - \frac{\left[ a+k^{\gamma}\right] }{n\left[ a+<k^{\gamma}>\right] }\right\rbrace + P\left( k-1,t,\gamma\right) \frac{\left[ a+\left( k-1\right) ^{\gamma}\right] }{n\left[ a+<k^{\gamma}>\right]},
\end{equation}
\end{widetext}
with $ P\left( k,t\right) $ satisfying the initial condition, $ P\left( k,0\right)= \delta_{k,0} $.
In the equation above, at time step $ t+1 $, the second term describes the average loss of the number of vertices of degree $ k $ because of a stub attaching to a vertex of degree $ k $ while the third term describes the gain in the number of vertices of degree $ k $ because of the attachment of a stub to a vertex of degree $ k-1 $. 
Since the maximum degree of any vertex at time $ t $ cannot exceed $ t $, $ P\left( k,t\right)= 0 $ for $ k > t $.
\par
It is difficult to solve the master equation exactly for all $ \gamma $.  However in the $ n \longrightarrow \infty $ limit, the equation can be solved exactly for $ \gamma = 1 $, the linear preferential attachment model, where it has been argued networks exhibit power law behaviour for large $ k $.  For $ \gamma < 1 $, even if we cannot derive the exact form of the distribution, we will deduce the leading large $ k $ behaviour of the degree distribution in the $ n \longrightarrow \infty $ limit.  We first recast the master equation in a form that makes it easier to get the leading terms in an $ 1/n $ expansion of the degree distribution.  To do that, notice that it is easy to get the form of the solution for $ k=0 $ by induction.  It is given by 
\begin{equation}
P\left( 0,t,\gamma\right) = \prod\limits_{j=0}^{t-1}\left\lbrace 1 - \frac{\left( a + \delta_{\gamma,0}\right)}{n\left[ a + <k^{\gamma}>\left( j\right) \right]  }\right\rbrace.
\end{equation}
The solution for the first few $ k $ and $ t $ values suggest trying a general solution of the form
\begin{equation}\label{ansatz}
P\left( k,t,\gamma\right) = \prod\limits_{i=0}^{k-1}\left( a + i^{\gamma}\right)\prod\limits_{j=0}^{t-1}\left\lbrace 1 - \frac{\left( a + k^{\gamma}\right) }{n\left[ a + <k^{\gamma}>\left( j\right) \right]}\right\rbrace  F\left( k,t,\gamma\right)  \qquad\mbox{for}\quad k \ge 1,
\end{equation}
where $ F\left( k,t,\gamma\right) $ is to be determined. Substituting this form for the solution in the master equation then gives
\begin{equation}\label{Fequation}
F\left( k,t+1,\gamma\right) = F\left( k,t,\gamma\right) + \frac{F\left( k-1,t,\gamma\right)}{n\left[ a + <k^{\gamma}>\left( t\right) \right]}L\left( k,t,\gamma\right)  \quad\mbox{where}
\end{equation}

\begin{equation}
L\left( k,t,\gamma\right) = \frac{\prod\limits_{l=0}^{t-1}\left\lbrace 1 - \frac{\left( a + \left( k-1\right)^{\gamma}\right)}{n\left[ a + <k^{\gamma}>\left( l\right)  \right]}\right\rbrace}{\prod\limits_{m=0}^{t}\left\lbrace 1 - \frac{\left( a + k^{\gamma}\right)}{n\left[ a + <k^{\gamma}>\left( m\right)\right]}\right\rbrace}.
\end{equation}
In Eq.(\ref{Fequation}), the second term explicitly is of $ O\left( 1/n\right)$, and so we can approximate and keep only the leading term in $ L\left( k,t,\gamma\right)$ in the limit $ n\longrightarrow\infty $. In the large $ n $ limit, the products in the numerator and denominator can be approximated by exponentials of sums giving
\begin{equation}\label{largenapprox}
L\left( k,t,\gamma\right)\approx e^{\frac{\left( a + k^{\gamma}\right)}{n\left[ a + <k^{\gamma}>\left( t\right)\right]}} \times e^{\frac{\left[ k^{\gamma}-\left( k-1\right)^{\gamma}\right]}{n}\sum\limits_{i=0}^{t-1}\frac{1}{\left[ a + <k^{\gamma}>\left( i\right)\right]}}.
\end{equation}
Because the first exponential in $ L\left( k,t,\gamma\right)$ introduces higher order $ 1/n $ terms it can be set to $1$.  In the large $ n $ limit we then get for $ F\left( k,t,\gamma\right)$ 
\begin{widetext}
\begin{equation}\label{basicFequation}
F\left( k,t+1,\gamma\right)= F\left( k,t,\gamma\right)+ \frac{F\left( k-1,t,\gamma\right)}{n\left[ a + <k^{\gamma}>\left( t\right) \right]}\, e^{\frac{\left[ k^{\gamma}-\left( k-1\right)^{\gamma}\right]}{n}\sum\limits_{i=0}^{t-1}\frac{1}{\left[ a + <k^{\gamma}>\left( i\right)\right]}}.
\end{equation}
\end{widetext}
We next look at two special cases of $ \gamma $ for solving the equation for the distribution $ P\left( k,t,\gamma \right)$.
\section*{Preferential Linear attachment}
In the preferential linear attachment model, $ \gamma=1 $ in the formulas for $ p_{k}\left( \gamma\right)$ in Eq. (\ref{probk>1})  and Eq. (\ref{probk0}) . Setting $ \gamma=1 $ in Eq. (\ref{ansatz}), and using the same approximation for large $ n $ we used in Eq. (\ref{largenapprox}), and using the fact that $ <k> = t/n $, the distribution function then becomes
\begin{equation}
P\left( k,t,1\right) = \frac{\Gamma\left( a + k\right)}{\Gamma\left( a\right)\Gamma\left( k + 1\right)}\, e^{-\left( a + k\right)  \sum\limits_{j=0}^{t-1}\frac{1}{\left[ na + j\right]}} \times F\left( k,t,1\right),
\end{equation}
and $ F $ is now given by
\begin{equation}
F\left( k,t + 1,1\right)= F\left( k,t,1\right) + \frac{F\left( k-1,t,1\right)}{\left[ na + t\right]}\, e^{^{\sum\limits_{i=0}^{t-1}\frac{1}{\left[ na + i\right]}}}.
\end{equation}
For large $ n $,
\begin{equation}
\sum\limits_{i=0}^{t-1}\frac{1}{\left[ na + i\right]} \approx \left\lbrace \Psi\left( na + t\right)- \Psi\left( n\right)\right\rbrace \approx \ln\left\lbrace  \frac{na + t}{na}\right\rbrace \quad\mbox{where}
\end{equation}
$ \Psi $ is the Psi function \cite{nisthandbook}.
The equation for $ F $ now takes the simple form
\begin{equation}
F\left( k,t+1,1\right) = F\left( k,t\right) + \frac{F\left( k-1,t,1\right)}{na},
\end{equation}
which can be solved using generating functions \cite{generatingfunctionology}. Defining the generating function of $ F(k,t,1) $ in the usual way as $ G(k,z,1)=\sum\limits_{t=0}^{\infty}F\left( k,t,1\right) z^{t} $ with $ \mid z \mid \leq 1 $, and $ G(k,0,1) = \delta_{k,0} $, it can be shown that $ G(k,z,1) $ satisfies the equation
\begin{equation}
G\left( k,z\right) = \left( \frac{z}{na}\right)^{k}\frac{1}{\left( 1-z\right) ^{k+1}}.
\end{equation}
The coefficient of $ z^{t} $ then gives us 
\begin{equation}\label{gammaonesolution}
F\left( k,t,1\right) = \frac{1}{\left( na\right) ^{k}}\frac{t!}{k!\left( t-k\right)!}.
\end{equation}
The degree distribution function in the $ n\longrightarrow\infty $ limit becomes
\begin{equation}
P\left( k,t,\gamma=1\right) = \frac{\Gamma\left( a+k\right)}{\Gamma\left( a\right) \left( na\right)^{k}}\frac{t!}{k!\left( t-k\right)!}\left[ \frac{na}{na+t}\right]^{a+k}.
\end{equation}
Since we are mainly interested in the $ t\longrightarrow\infty $ limit, such that $ t/n = \bar{k} $ is finite, we get
\begin{equation}
P\left( k,t,\gamma=1\right) = \frac{\Gamma\left( a+k\right)}{\Gamma\left( a\right) \Gamma\left( k+1\right)}\frac{1}{\left[ 1+\bar{k}/a\right]^{a}}\left[ \frac{\bar{k}/a}{1 + \bar{k}/a}\right]^{k}.
\end{equation}
Importantly, in the large $ k $ limit, as $ k\longrightarrow\infty $,
\begin{equation}
P\left( k,t,\gamma=1\right) \sim k^{a-1} e^{-k\ln\left( 1 + a/\bar{k}\right)},
\end{equation}
and the leading $ k $ dependence falls off exponentially rather than as a power law.
 
\section*{Non-linear attachment}
We will show that the leading large $ k $ dependence falls off exponentially for arbitrary $ 0 < \gamma < 1 $ even though we have not been able to get an exact form for the degree distribution. We start with the basic equation for $ F\left( k,t,\gamma\right) $ given by Eq. (\ref{basicFequation}) .  We already saw that for $ \gamma = 1 $, $ F $ is given by Eq. (\ref{gammaonesolution}) .  In the opposite limit of $ \gamma = 0, < k^{\gamma}> = 1 $,  and the equation for $ F $ takes the form
\begin{equation}
F\left( k,t+1,0\right) = F\left( k,t,0\right) + \frac{F\left( k-1,t\right)}{n\left( a + 1\right)},
\end{equation}
which can again be solved easily by the generating function method to give
\begin{equation}
F\left( k,t,0\right) = \frac{1}{n^{k}\left( a + 1\right)^{k}}\frac{t!}{k!\left( t-k\right)!}.
\end{equation}
The solution for the degree distribution for arbitrary $ \gamma $, and large $ t $ now takes the form
\begin{equation}
P\left( k,t,\gamma \right) = \frac{\bar{k}^{k}}{k!} \frac{1}{f\left( k,\gamma\right)} \prod\limits_{i=0}^{k-1} \left( a + i^{\gamma}\right) \,e^{-\frac{\left( a+k^{\gamma}\right)}{n}\sum\limits_{j=0}^{t-1}\frac{1}{\left[ a+<k^{\gamma}>\left( j\right)\right]}} ,
\end{equation}
where $ f\left( k,\gamma\right) $, though an unknown function, goes from $ f\left( k,\gamma = 0\right)  = \left( a+1\right) ^{k} $ to $ f\left( k,\gamma = 1\right) = a^{k} $. Since $ \prod\limits_{i=0}^{k-1}\left( a+i^{\gamma}\right) < \left( k-1\right)^{k} $, for large $ k $, the degree distribution becomes 
\begin{equation}
P\left( k,t,\gamma\right) < \frac{\bar{k}^{k}}{\sqrt{k}f\left( k,\gamma\right)} \,e^{-k} \,e^{-\frac{\left( a+k^{\gamma}\right)  }{n}\sum\limits_{j=0}^{t-1}\frac{1}{\left[ a+<k^{\gamma}>\left( j\right)\right]}}.
\end{equation}
We expect that $ f\left( k,\gamma\right) $ is a smooth well-behaved function of $ \gamma $ between the two limits $ \gamma = 0 $ and $ \gamma = 1 $ in which case $ f \approx a^{k} $.  Since the summation in the exponential is a function only of $ t $, we can conclude that the leading behaviour of $ P\left( k,t,\gamma\right) $ for large $ k $ is still a simple exponential.
\par
In conclusion, we have shown that in models of preferential attachment, the resulting degree distribution for large degrees falls off exponentially when the model starts with a fixed number of vertices and adds edges so that they attach themselves to the vertices with a probability proportional to a power of the degree of the vertex.  This is in contrast to preferential attachment models that are based on citation networks, where the network is constructed by growing both the vertices and the edges and where it has been shown that the resulting degree distribution for large degree is either scale free or a stretched exponential.  Since it is known that in these models earlier vertices acquire excess edges, we conclude that preferential attachment alone is not a sufficient condition for generating scale free networks.

\bibliography{randomgraph}

\end{document}